\documentclass[12pt,a4]{article}\input{epsf.tex}
\begin{document}
\begin{center}
{\bf LEVEL DENSITY AND RADIATIVE STRENGTH FUNCTIONS OF DIPOLE
$\gamma$-TRANSITIONS IN $^{139}$Ba AND $^{165}$Dy}\\
\end{center}
\begin{center}
{\bf V.A. Khitrov, A.M. Sukhovoj}\\
{\it Frank Laboratory of Neutron Physics, Joint Institute
for Nuclear Physics, 141980, Dubna, Russia}\\

{\bf Pham Dinh Khang}\\ 
{\it National University of Hanoi}\\
{\bf Vuong Huu Tan,  Nguyen Xuan Hai}\\
{\it Vietnam Atomic Energy Commission}\\ \end{center}

\begin{abstract}
Level density and radiative strength functions which allow precise reproduction
 of the two-step cascade intensity, gamma width of compound state and cascade
 population of levels up to excitation energy of about 3.5 MeV were determined
using experimental data on the $(n,2\gamma)$ reaction.
Level density in these nuclei (like in other even-odd nuclei studied earlier)
in wide excitation energy interval is considerably less than that predicted by
 Fermi-gas model. Enhancement of the radiative strength functions, caused by
strong correlations between cascade gamma-decay parameters, most probably,
relates with the change in ratio between the quasi-particle and collective
components of the wave functions of the cascade intermediate levels in the 
region of most strong change in their density.
\end{abstract}
\section*{Introduction
} \hspace*{14pt}

The authors [1] have studied Cooper pairs breaking process by comparison of
level density in threes of isotopes with the neutron number $N-1,~N,~N+1$.
The confidence level in determination of nuclear specific heat
is fully stipulated by reliability of the observed level density.
This nuclear parameter obtained from intensities of two-step cascades has
 considerably higher reliability than that obtained within known methods 
 due to unsuccessful relation
between the experimental spectra and desired parameters of the gamma-decay
process. The data needed for the analysis [1] can be obtained now, 
for example, for three compound nuclei $^{137,138,139}$Ba and $^{163,164,165}Dy$.

Analysis method [2] of a bulk of the information on the two-step cascades,
developed in Dubna without using any the model notions of level density 
$\rho$ and radiative strength functions $k$, showed that these data cannot 
be reproduced without the ``step-like" structure in level density and 
corresponding deviations of $k$ from the simple model dependencies.
For the fist time, the possibility of the ``step-like" structures in the level 
density and corresponding thermodynamic characteristics of a nucleus was
 pointed out in [3].

\section{Analysis}
\hspace*{14pt}
Using our experimental data on cascade $\gamma$-transitions from
the $^{139}$Ba$(n,2\gamma)$  [4] and  $^{164}$Dy$(n,2\gamma)$ [5] reactions
we determined [6] the dependence of
the two-step cascade intensity $I_{\gamma\gamma}(E_1)$ on the primary
transition energy $E_1$ for these nuclei  (it is shown in Fig.~1), and also level density
$\rho$ and radiative strength
functions $k=\Gamma_{\lambda i}/(E_\gamma^3A^{2/3}D_\lambda)$ which allow us
to reproduce $I_{\gamma\gamma}(E_1)$ with practically zero 
deviation from the experiment for both nuclei.
The most considerable errors of procedure [6] of
decomposition of the experimental spectrum into two components corresponding
to solely primary transitions and solely secondary transitions - leads only 
to re-distribution of cascade intensities between the
different intervals of the primary transition energies.

In the first approach, magnitude of this error is inversely proportional to 
the number of cascades registered in the experiment. It increases also with 
increasing a number of background events.

Although efficiency of the spectrometer used in the experiment [4,5] was small
enough, rather specific form of energy dependence of the cascade intensity
(considerable concentration of $I_{\gamma\gamma}$ at the energy of their
intermediate levels $E_i \simeq 0.5B_n$ for $^{165}$Dy) decreases the influence of systematical
 uncertainties of the procedure [6] for the primary transitions of cascades with 
 the energy $E_1 \leq 2$ MeV. In practice, this can increase cascade intensity 
 for $E_1 \ge 2$ MeV not more than by 20-30\% and, respectively, decrease it 
 in the same measure in the region $E_1 \ge 3.7$ MeV.

In case of  $^{165}$Dy, the $\rho$ and $k$ data obtained from analysis [2] are slightly
 misrepresented. But the use of the method [6] does not change the total cascade
 intensity in the $2 < E_1 < 3.7$ MeV interval. Namely, considerable exceeding
 of experimental intensity at high enough nuclear excitation $E_{ex}=B_n-E_1$ above the
 calculation in the framework Fermi-gas [7] model of level density or any
 other notions (which provide exponential energy dependence of level
 density in the interval from 1-2 MeV and $B_n$) gives ``step-like" structure
 in energy dependence of level density. As a consequence, this systematical
 error leads, in practice, to insignificant variation in values of the desired
 parameters $\rho$ and $k$. 
Just this circumstance allow us to get the data on the level density and radiative
 strength functions with relatively small systematic errors.

Additional error can result from systematic errors of intensities of the cascade
 high-energy primary gamma-transitions which are used for normalization of
 $I_{\gamma\gamma}$. Intensities [5] used for this aim are about 25\% less than
 the modern values [9]. So, one can summarize that all the known ordinary
systematic
 errors cannot decrease the presented below experimental level density at
 least at the excitation energy $E_{ex}< 3$ MeV.
Such systematical error can not explain the difference between experimental and
calculated data in $^{139}$Ba even in principle also.

Level density and radiative strength functions of $E1$ and $M1$ cascade
transitions, which allow simultaneous reproduction of cascade intensity
$I_{\gamma\gamma}(E_1)$ (Fig.~1) and the mean value of the total radiative
width $\Gamma_{\lambda}=55$ and 57 meV [10] of neutron resonances in
$^{138}$Ba and $^{164}$Dy correspondly,
are shown in Figs.~2 and 3, respectively.

It was established experimentally in first time now [12] that the ratio
$k(E_\gamma,E_{ex})/k(E_\gamma, B_n)$ of
 strength functions for transitions with equal energy and multipolarity but
 depopulating levels with significantly different excitation energy strongly
 depends on $E_{ex}$. This must lead to significant discrepancy between the
 calculated and experimental single spectra of cascade gamma-decay of any
 nuclei and to noticeably less discrepancy in the case of intensities of
 two-step cascades to the most low-lying levels.

This is due to the sign-variable variation of energy dependence of
 strength functions $k(E_\gamma, E_{ex})$ of the secondary transitions
 with respect to $k(E_\gamma, B_n)$ for the primary transitions.
 Such variations influence corresponding $I_{\gamma\gamma}$ values
 only through the considerably less change in the total radiative
 strength of the cascade intermediate levels. But even in this case,
 the errors in parameters can exceed the width of the interval of
 probable values $\rho$ and $k$ providing precise description of the
 experimental cascade intensities.

The use of cascade population of the large enough set of cascade intermediate
 levels with rather high maximum excitation energy provides, to the first
 approach, accounting [12] for the dependence $k(E_\gamma, E_{ex})$.
 Corresponding results for population by cascades and primary transitions
 of such set of levels are shown in Fig.~4, the lower estimates of cascade
 populations summed over the 200 keV energy bins are shown in Fig.~5.
 The last data can be added to the set of experimental data using for extraction [2]
 of $\rho$ and $k$. This provides as low as possible error in determination of the
 parameters to be found.

As in other studied even-odd nuclei, level density in $^{139}$Ba $^{165}$Dy in the
 excitation interval from $\sim 1$ to $\sim 3$ MeV also is considerably less than that
 predicted according to [7]. Theoretical basis for qualitative explanation
 of such energy dependence
was obtained in [3]. In accordance with the main idea by A.V. Ignatyuk and
Yu.V. Sokolov, quasi-particle level density is the sum of level densities 
with 1, 3, 5... quasi-particles for even-odd nuclei. In the interval
 between the energies of breaking of corresponding Cooper pairs, level
 density changes very weakly, at least for some first broken pair.
 And energy of a nucleus, most
probably, is passed to excitation of its vibrations. The only correction
which is necessary to achieve well agreement between the experiment and
calculation [3] is the shift of the energy of appearance
of 3 quasi-particles to the higher value by about 1 MeV.
Enhancement of the radiative strength functions, caused by strong
correlations of $k$ and $\rho$, most probably, relates with the change in 
ratio between the quasi-particle and collective components of the wave 
functions of the cascade intermediate levels in the region of most strong
change in their density.

\section{Conclusion}\hspace*{14pt}

Method [2] does not allow one, even in principle, to get unique values of
the level density and radiative strength functions even in that case when
experimental cascade intensities do not contain systematical and statistical errors.
Modeling of this situation shows that their asymptotical uncertainty
for the available experimental data cannot be less than $\approx 20\%$.
This results from:

(a)~exceeding of number of the determined parameters over the number
of experimental points, and

(b)~specific form of the functional dependence of $I_{\gamma\gamma}$ on
the desired parameters $\rho$ and $k$.

But even in this case the determined parameters should be considered as the
most reliable among the analogous results, first of all, due to model-free
method of their determination.

\begin{flushleft}
{\large\bf References}
\end{flushleft}
\begin{flushleft}

1. K. Kaneko and M. Hasegawa, Nucl. Phys. A, {\bf 740} (2004) 95.\\
2. E.V. Vasilieva, V.A. Khitrov, A.M. Sukhovoj, Phys. of Part. and
Nucl., \\\hspace*{14pt} {\bf 31(2)} (2000) 170.\\
\hspace*{14pt}
E.V. Vasilieva, A.M. Sukhovoj, V.A. Khitrov,
Phys. of Atomic Nuclei.\\\hspace*{14pt} {\bf 64(2)} (2001) 153.\\
3. A.V. Ignatyuk, Yu.V. Sokolov, Yad. Fiz., (1974) {\bf 19} 1229\\
4. V.A.Bondarenko et al., Sov.J.Nucl.Phys. {\bf 54}(1991) 545.\\
5. Yu.P.Popov, A.M.Sukhovoi, V.A.khitrov, Yu.S.Yazvitsky, \\\hspace*{14pt}
Izv.Akad.Nauk SSSR, Ser.Fiz. {\bf 48} (1984) 891.\\
6. S.T. Boneva et. al.//Nucl. Phys. {\bf A589} (1995) 293.\\
7. W. Dilg, W. Schantl, H. Vonach and M. Uhl, Nucl. Phys. {\bf A217}
(1973) 269.\\
8. http://www-nds.iaea.org/pgaa/egaf.html\\ \hspace*{14pt}
G.L. Molnar et al., App. Rad. Isot. (2000) {\bf 53} 527\\
9. Axel P., Phys. Rev. 1962, V. 126(2), P. 671.\\ 
 \hspace*{14pt}Blatt J. M., Weisskopf V. F. Theoretical Nuclear Physics. New York (1952).\\
10. S.F. Mughabghab, Neutron Cross Sections. V. 1. Part B. N.Y. Academic
Press, \\\hspace*{14pt}(1984)\\
11. S.G. Kadmenskij, V.P. Markushev, V.I. Furman, Sov. J. Nucl. Phys. {\bf 37}
(1983) 165.\\
12. V.A. Bondarenko et all, In: XII International Seminar on Interaction
of Neutrons \\\hspace*{14pt}with Nuclei,  Dubna, May 2004,
E3-2004-169, Dubna, 2004, p. 38.\\
\end{flushleft}
\newpage
\begin{figure}[htbp]
\begin{center}
\leavevmode
\epsfxsize=14cm
\epsfbox{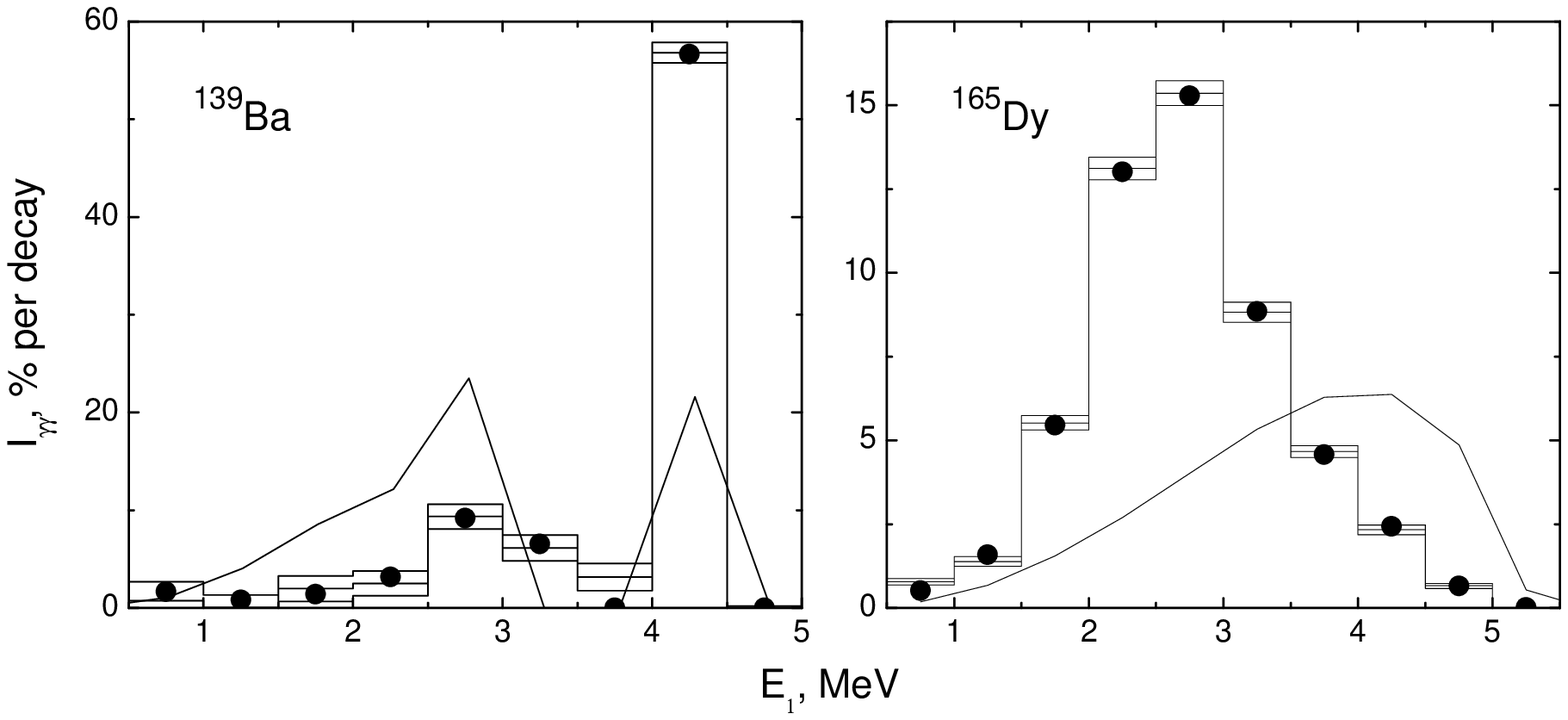}
\end{center}
\hspace{-0.8cm}\vspace{-14.cm}

{\bf Fig.~1} ~Histogram is the total experimental intensities
 of two-step cascades (summed in energy bins of 500 keV) with ordinary
statistical errors as a function of the primary transition energy. 
Line is the calculation in frame of models [7,9]. 
Points are the typical fit by the most probable values $\rho$ and $k$.
\end{figure}
 \begin{figure}
\begin{center}
\vspace{-2cm}
\leavevmode
\epsfxsize=14cm
\epsfbox{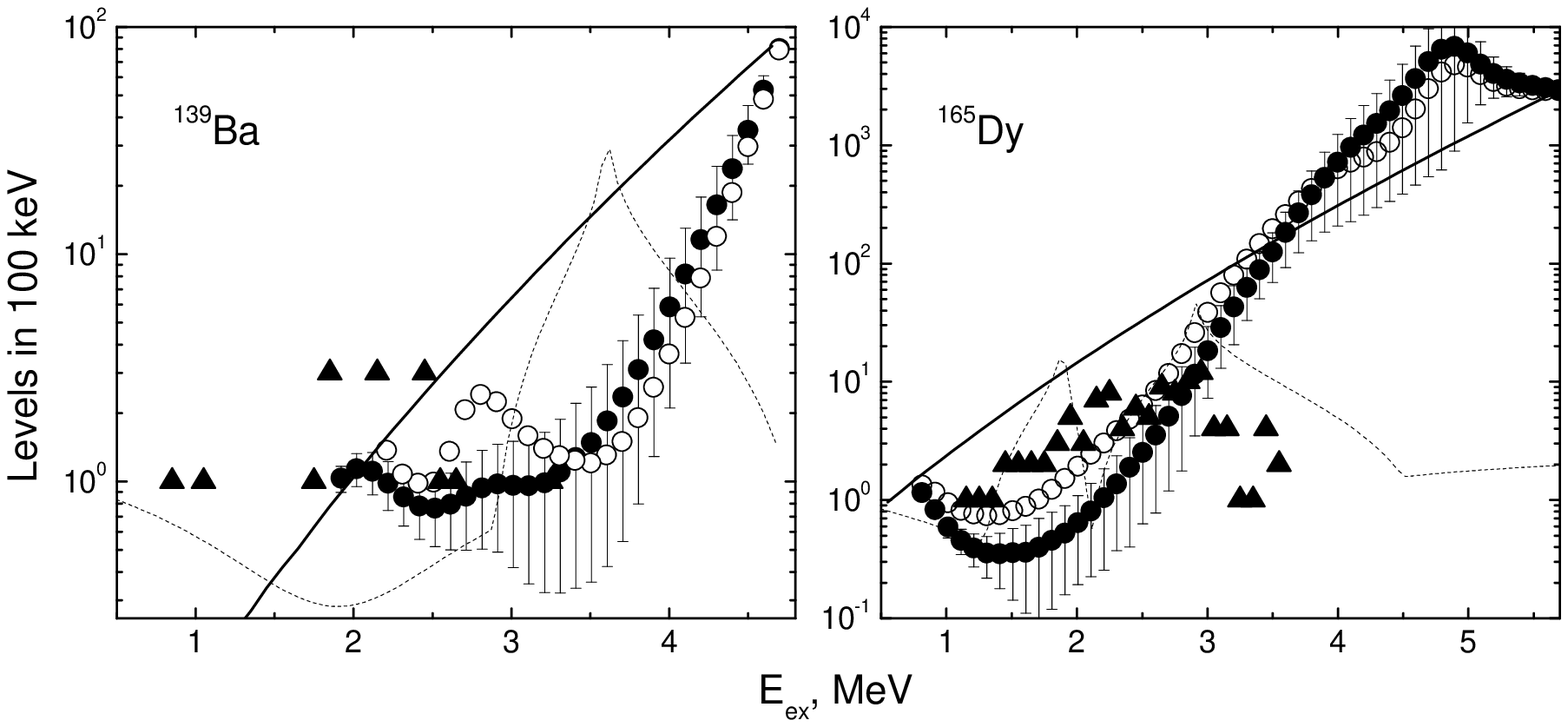}
\end{center}
\hspace{-0.8cm}\vspace{-15.cm}

{\bf Fig.~2} ~Full circles are the expected number of levels for both parities
 and spins 1/2, 3/2 in case of different functional dependence of strength
 functions for primary and secondary cascade transitions. Open circles are the
 same for equal functional dependence of strength functions for primary and
 secondary cascade transitions. Dash lines shows values of function $h$ [12]
 for excitation energy $E_{ex}=B_n-E_1$. Lines represents predictions
 according to model [7]. Triangles are the number of obtained in [4,5] two-step
 cascades intermediate levels.
\end{figure}
\newpage
\begin{figure}
\begin{center}\leavevmode
\vspace{-2cm}
\epsfxsize=14cm
\epsfbox{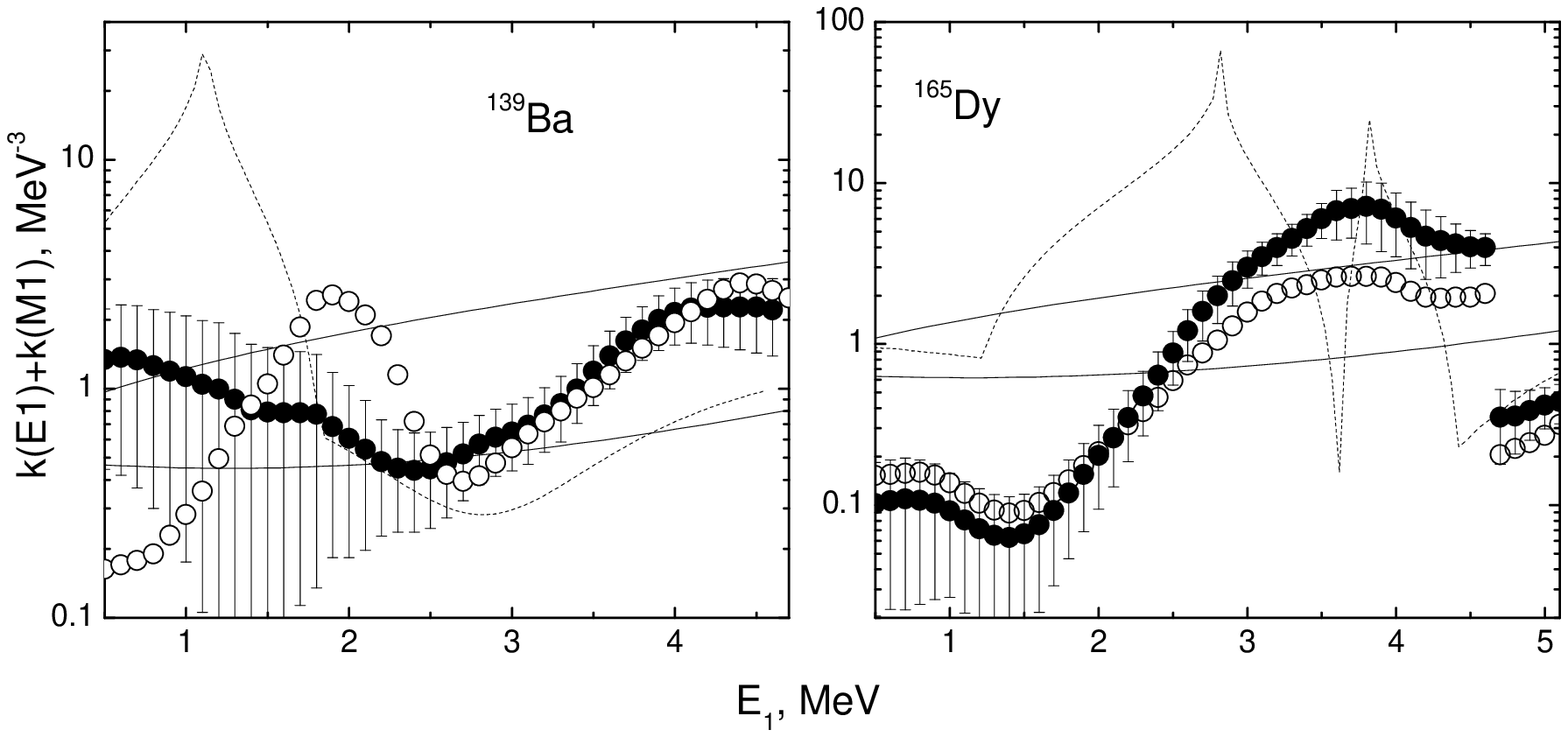}\end{center}
\hspace{-0.8cm}\vspace{-12.cm}

{\bf Fig.~3} ~The sums of radiative strength functions of the cascade primary
 dipole
transitions providing reproduction of cascade intensities with the considered 
difference of their values with strength functions of secondary transitions 
(multiplied by $10^9$). Open circles are the same for one equal energy
 dependence of strength functions for primary and secondary cascade transitions.
Dash lines shows values of function $h$ for excitation energy $B_n-E_1$.
Lines are the models [9,11] predictions with $k(M1)=const$.
\end{figure}

\begin{figure}[htbp]
\vspace{-2cm}
\leavevmode
\epsfxsize=14cm

\epsfbox{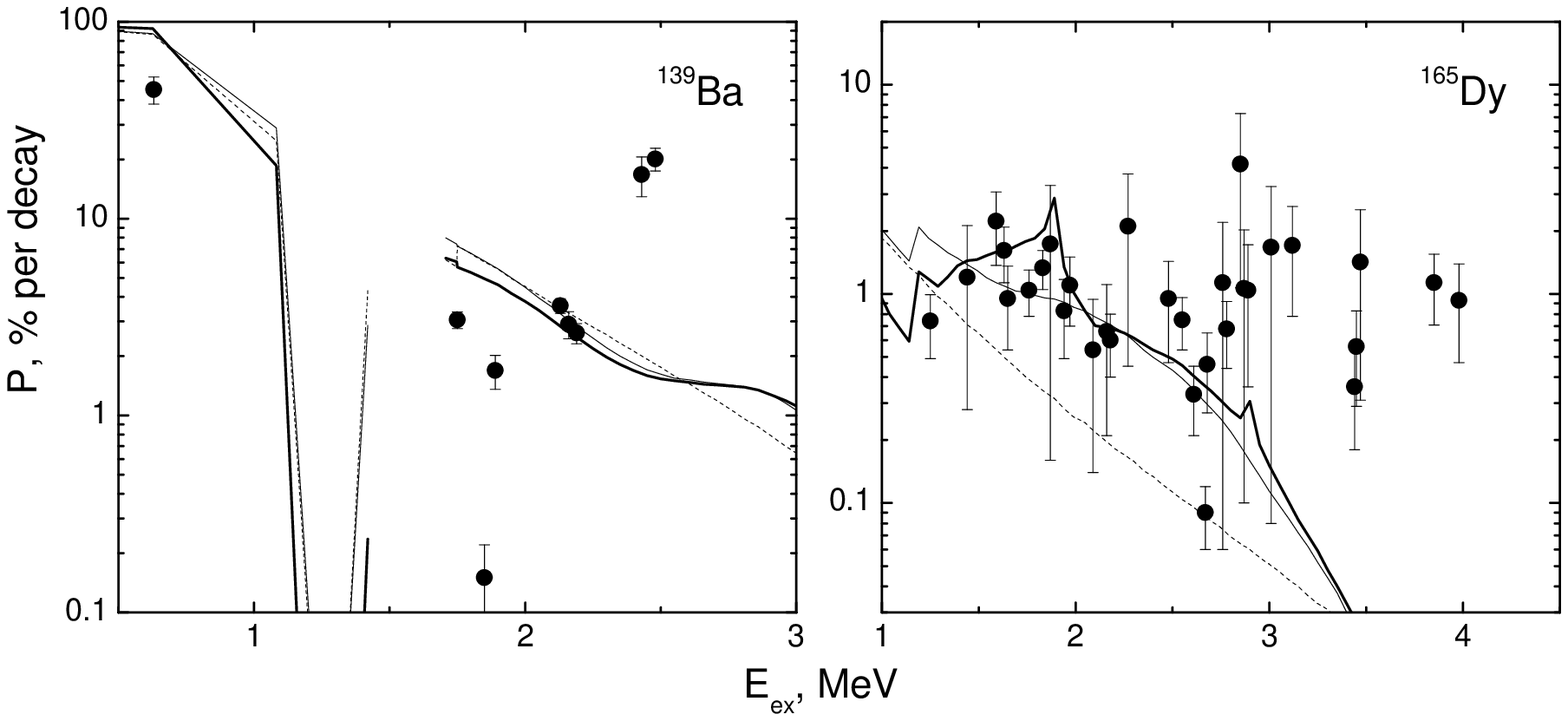}
\vspace{-14.cm}

{\bf Fig. 4.}  ~The total population of intermediate levels of two-step
 cascades (points with bars), dashed
line represents calculation within models [7,9].
Thin line shows results of calculation using data [2].
Thick line shows results of 
calculation using level density [2], and corresponding strength functions
of secondary transitions are multiplied by function $h$ determined within method 
 [12].
\end{figure}

\begin{figure}[htbp]
\vspace{-2cm}
\leavevmode
\epsfxsize=14cm

\epsfbox{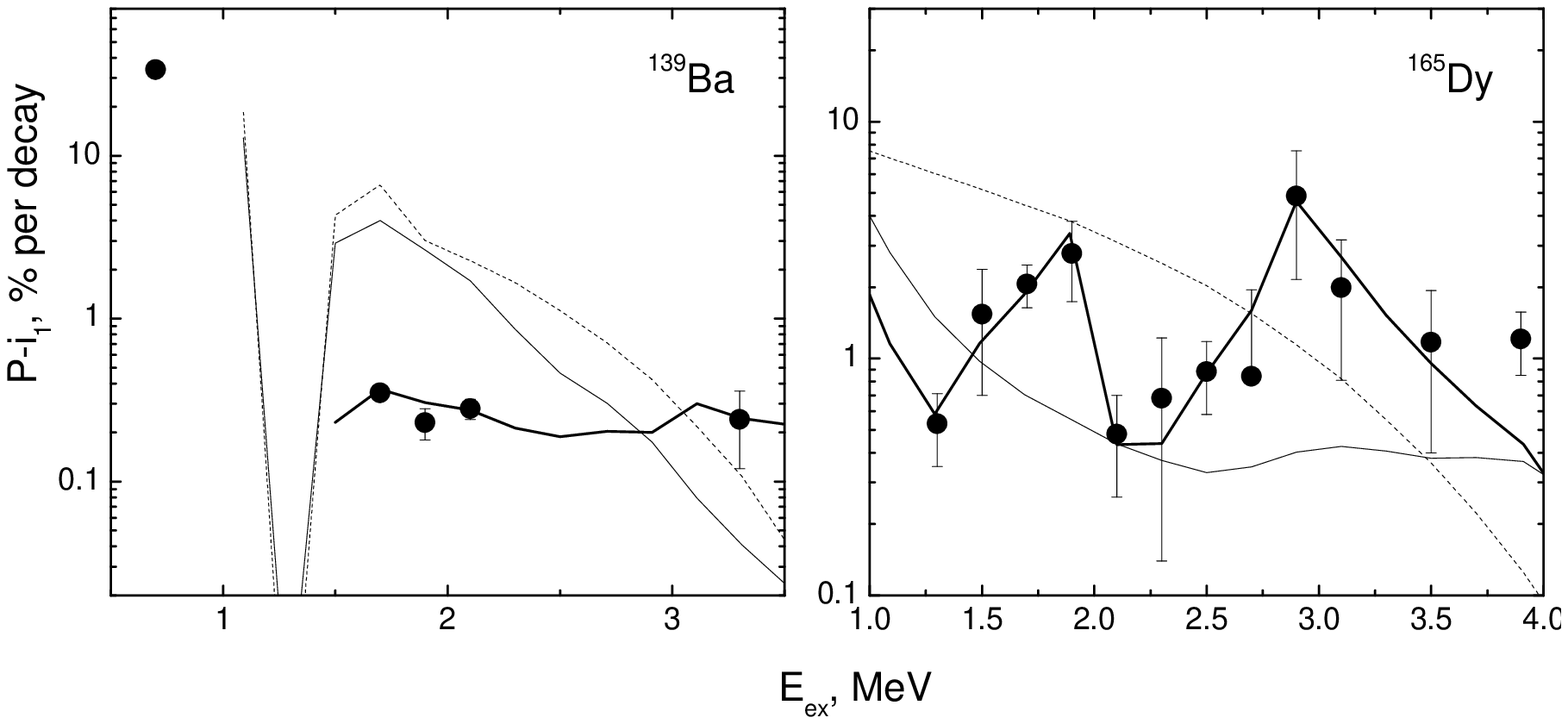}
\vspace{-14.cm}

{\bf Fig. 5.} ~The same as in Fig.~4 for sum of the cascade population only for
levels in the 200 keV energy bins.
\end{figure}
\end{document}